\documentclass[twocolumn,prl,floats]{revtex4}

\usepackage{graphics}

\begin{document}
\title{Quantum correlated twin atomic beams via photo-dissociation of a molecular \\
Bose-Einstein condensate}
\author{\vspace{-0.3cm}K. V. Kheruntsyan and P. D. Drummond}
\affiliation{{\it Department of Physics, The University of Queensland, Brisbane,
Queensland 4072, Australia}}
\date{7 September 2001}
\begin{abstract}
We study the process of photo-dissociation of a molecular Bose-Einstein
condensate as a potential source of strongly correlated twin atomic beams.
We show that the two beams can possess nearly perfect quantum squeezing in
their relative number.

\end{abstract}
\maketitle
The successful production of Bose-Einstein condensates (BEC) has led to
measurements indicating that BECs have coherence properties similar to
lasers. This suggests that the next stage in obtaining precision
measurements in atom optics is the production of atom beams with
sub-Poissonian atom statistics -- as has been widely demonstrated in
photonics applications. Indeed, there is already some indirect evidence of
atom-number squeezing, but experiments and theoretical proposals to date
have focussed on trapped condensates and phase sensitive measurements \cite
{Squeezed States in BEC,Moelmer2001,Burnett2001}. A possible route towards
more robust phase-insensitive applications of atomic squeezing is to produce
quantum correlated atom laser beams in which the correlations are directly
built in the statistics of the particle numbers, as in parametric
down-conversion in quantum optics \cite{Heidmann}. Quantum correlated or
entangled photon pairs from parametric down-conversion are one of the most
powerful resources of quantum optics, and our motivation here originates
from the intriguing prospect of possessing a {\em matter-wave} analog of
this resource. As well as the immediate possibility of improved atomic
interferometry, an exciting prospect would be the development of new tests
of quantum measurement theory for massive particles with space-like
separations, since all previous tests using down-conversion methods were
restricted to massless photons.

In this paper we propose a robust scheme for achieving strong quantum
correlations between two counter-propagating atomic beams, relying on the
process of photo-dissociation of a molecular Bose-Einstein condensate.
Experiments towards production of molecular condensates \cite
{WFHRH2000,Towards molecular BEC}, together with a number of theoretical
studies of coupled atomic-molecular BEC systems are the subject of much
intense activity at present \cite{Superchemistry,AM-BEC via PA,AM-BEC via
Feshbach}. We anticipate that the formation of a molecular BEC is a matter
of time, and consider it as the starting point for a twin atomic beam
experiment with relative atom number fluctuations reduced below the level
predicted by either thermal or coherent (Poissonian) statistics. The method
automatically yields two counter-propagating beams through momentum
conservation, and is robust against changes in mode structure, coupling
constants, or even absorptive losses, provided they are small.

An important feature of our proposal is that it does not rely on atomic
interferometry or local oscillators to generate the resulting correlations,
which makes it more practical than recent related proposals 
\cite{Moelmer2001,Burnett2001}. In addition, the present scheme is not
susceptible to phase noise from self-phase modulation. In the
following analysis, we first analyze a simplified theory of a uniform,
non-depleted molecular condensate, then include the effects of molecular
trapping, depletion, phase diffusion and atomic losses.

The quantum field theory effective Hamiltonian for the atomic ($\hat{\Psi}
_{1}$) and molecular ($\hat{\Psi}_{2}$) fields, taken for simplicity to be
confined to one space dimension, is: 
\begin{eqnarray}
\hat{H} &=&\hat{H}_{kin}+\int dx\left\{ \sum_{i}V_{i}(x)\hat{\Psi}_{i}^{\dag
}\hat{\Psi}_{i}^{{}}+\sum_{i\geq j}U_{ij}\hat{\Psi}_{i}^{\dag }\hat{\Psi}
_{j}^{\dag }\hat{\Psi}_{j}\hat{\Psi}_{i}\right.   \nonumber \\
&&\left. -i\frac{\chi (t)}{2}\left[ e^{i\omega t}\hat{\Psi}_{2}^{\dagger }
\hat{\Psi}_{1}^{2}-e^{-i\omega t}\hat{\Psi}_{2}\hat{\Psi}_{1}^{\dagger \,2}
\right] \right\} ,
\end{eqnarray}
with the commutatotion relation $[\hat{\Psi}_{i}(x,t),\hat{\Psi}_{j}^{\dagger
}(x^{\prime },t)]=\delta _{ij}\delta (x-x^{\prime })$. Here $\hat{H}_{kin}$
stands for the usual kinetic energy term, $V_{i}(x)$ is the trap potential
(including internal energies), $U_{11}\simeq 4\pi \hbar a_{1}/(Am_{1})$ is
the atom-atom coupling constant in one dimension, where $m_{1}$ is the mass, 
$a$ is the three-dimensional $S$-wave scattering length, and $A$ is the
confinement area in the transverse direction, with similar results for the
molecule-molecule and molecule-atom interactions. The term
proportional to $\chi (t)$ describes a coherent process of molecule-atom
conversion via either one or two-photon (Raman) photo-dissociation, where $
\chi (t)$ is the coupling constant related to the transition matrix
element(s) and the amplitude(s) of the photo-dissociation laser(s) which
have an overall frequency (difference) $\omega $.

The time dependence of $\chi (t)$ controls the duration of dissociation
process and will be set to $\chi (t)=\chi _{0}\theta (t)\theta (t_{1}-t)$,
where $\theta $ is a step function. This means that the dissociation can be
stopped after a short time interval $t_{1}$ followed by free evolution of
the atomic field. Starting from a pure molecular condensate, the molecular
field can be initially described semiclassically, via its initial coherent
amplitude $\Psi _{2}(x,0)=\langle \hat{\Psi}_{2}(x,0)\rangle $. We assume $
n_{2}(x)=|\Psi _{2}(x,0)|^{2}$ is the initial molecular BEC density in a
harmonic trap in the Thomas-Fermi limit, with the axial half-length $x_0$.

The atoms are assumed to be untrapped longitudinally (they may be in an $m=0$
magnetic sublevel) yet confined transversely (they may be in a transverse
optical trap), so that the atomic field can effectively be treated as a free
one-dimensional field, initially in a vacuum state. One-dimensional trapping
of condensates has been achieved experimentally \cite{1D_BEC}, so this
is not unrealistic. In what follows, we will neglect the atom-atom
collisions for simplicity, owing to the fact that we restrict ourselves to
short interaction times during which the atomic density does not grow to
high values, so that the self-interaction term proportional to $U_{11}$ 
can be neglected. 
In addition, we choose the absolute value of the detuning $|\Delta |$ to be always
non-zero and much larger than the magnitude of $U_{12}\langle \hat{\Psi}
_{2}^{\dagger }\hat{\Psi}_{2}^{{}}\rangle $ so that the atom-molecule $S$
-wave scattering is negligible too. The detuning $\Delta \equiv V_{1}(0)-[V_{2}(x_{0})+\omega ]/2$,
where $V_{2}(x_{0})=V_{2}(0)+U_{22}n_{2}(0)$, is proportional to the energy
mismatch between the atomic and molecular fields.

The system has a direct analogy with travelling-wave parametric
down-conversion in non-linear optics \cite{Yariv}. Here the role of $\chi
_{0}$ is played by a non-linear crystal with second-order susceptibility $
\chi ^{(2)}$, and a finite interaction time is analogous to the crystal
length $L=vt_{1}$in the direction of propagation, where $v$ is the
group velocity of the fundamental (higher frequency) beam. The detuning $\Delta$
is analogous to the optical phase mismatch, 
while the atomic kinetic energy is analogous to dispersion.

To proceed with the analysis we introduce a characteristic length scale $
d_{0}$ and time scale $t_{0}=2m_{1}d_{0}^{2}/\hbar $. Next, transform to
dimensionless fields, in rotating frames: 
\begin{eqnarray}
\hat{\psi}_{1}(\xi ,\tau ) &=&\sqrt{d_{0}}\hat{\Psi}
_{1}(x,t)e^{i[V_{2}(x_{0})+\omega ]t/2},  \nonumber \\
\hat{\psi}_{2}(\xi ,\tau ) &=&\sqrt{d_{0}}\hat{\Psi}
_{2}(x,t)e^{iV_{2}(x_{0})t},
\end{eqnarray}
where $\xi =x/d_{0}$ and $\tau =t/t_{0}$ are the dimensionless coordinate
and time. We also introduce dimensionless detuning $\delta =\Delta t_{0}$
and coupling $\kappa (t)=\chi (t)t_{0}/\sqrt{d_{0}}$.

The Heisenberg equations of motion for the field operators, in dimensionless
form are: 
\begin{eqnarray}
\frac{\partial \hat{\psi}_{1}(\xi ,\tau )}{\partial \tau } &=&i\frac{
\partial ^{2}\hat{\psi}_{1}}{\partial \xi ^{2}}-i\delta \hat{\psi}
_{1}+\kappa \hat{\psi}_{2}\hat{\psi}_{1}^{\dag },  \nonumber \\
\frac{\partial \hat{\psi}_{2}(\xi ,\tau )}{\partial \tau } &=&\frac{i}{2}
\frac{\partial ^{2}\hat{\psi}}{\partial \xi ^{2}}-i\hat{v}_{2}(\xi )\hat{\psi
}_{2}-\frac{1}{2}\kappa \hat{\psi}_{1}^{2}.
\label{Heizneberg field-operator equations}
\end{eqnarray}
We have introduced an
effective potential $\hat{v}_{2}(\xi )=[V_{2}(\xi d_{0})-V_{2}(\xi
_{0}d_{0})]t_{0}+u\hat{\psi}_{2}^{\dag }\hat{\psi}_{2}$, where $
u=U_{22}t_{0}/d_{0}$, for notational simplicity.

To gain some insight into the underlying physics of correlated atomic beams,
we first consider an idealized and analytically solvable model corresponding
to an undepleted and uniform molecular condensate at density $n_{2}(0)$,
that fills the entire space from $-l/2$ to $l/2$, with periodic boundary
conditions at $-l/2$ and $l/2$. The atom-molecule coupling $\chi =\chi _{0}$
is assumed to be constant during the whole evolution time from $0$ to $\tau$. 
In this case the coherent amplitude of the molecular field can be absorbed
into an effective gain constant $g=\kappa _{0}\sqrt{n_{2}(0)d_{0}}$ (where $
\kappa _{0}=\chi _{0}t_{0}/\sqrt{d_{0}}$) which we assume without loss of
generality to be real and positive.

Solutions to the resulting linear set equations of motion for the atomic
field are easily found in momentum space, where we expand $\hat{\psi}
_{1}(\xi ,\tau )$ in terms of single-mode annihilation operators: $\hat{\psi}
_{1}(\xi ,\tau )=\sum_{q}\hat{a}_{q}(\tau )e^{iq\xi }/\sqrt{l}$, where $
q=d_{0}k$ is a dimensionless momentum. The single-mode operators $\hat{a}
_{q} $ satisfy the usual commutation relations $[\hat{a}_{q}(\tau ),\hat{a}
_{q^{\prime }}^{\dagger }(\tau )]=\delta _{qq^{\prime }}$. The corresponding
Heisenberg equations of motion have the following solution 
\begin{eqnarray}
\hat{a}_{q}(\tau ) &=&A_{q}(\tau )\hat{a}_{q}(0)+B_{q}(\tau )\hat{a}
_{-q}^{\dagger }(0),  \nonumber \\
\hat{a}_{-q}^{\dagger }(\tau ) &=&B_{q}(\tau )\hat{a}_{q}(0)+A_{q}^{\ast
}(\tau )\hat{a}_{-q}^{\dagger }(0).
\end{eqnarray}
where 
\begin{eqnarray}
A_{q}(\tau ) &=&\cosh \left( g_{q}\tau \right) -i\lambda _{q}\sinh \left(
g_{q}\tau \right) /g_{q}\,\,,  \nonumber \\
B_{q}(\tau ) &=&g\sinh \left( g_{q}\tau \right) /g_{q}\,\,,
\end{eqnarray}
with $\lambda _{q}\equiv q^{2}+\delta $, and $g_{q}=(g^{2}-\lambda
_{q}^{2})^{1/2}$. Solutions of this type to the classical counterpart of the
operator equations are well known in optics \cite{Yariv}, while in quantum
optics the operator equations in the context of squeezing of nonlinear
propagating fields were studied in \cite{Paramp}.

Knowledge of the initial state of the atomic field, which is the vacuum
state with $\hat{a}_{q}(0)\left| 0\right\rangle =0 $, allows us to calculate
any operator moments at time $\tau $. The parameter $g_{q} $ is the gain
coefficient; if real, it causes a growing correlated output for the momentum
component $q $, while if imaginary it leads to oscillations.

For example, the particle number distribution in momentum space is given by $
\left\langle \hat{a}_{q}^{\dagger }(\tau )\hat{a}_{q}(\tau )\right\rangle
=B_{q}^{2}(\tau )$. For $\delta <0$, the function $B_{q}^{2}$ has two
distinct global maxima located at $q$-values where $\lambda _{q}=0$. This
gives the two most probable momentum values $q_{0}=\pm \sqrt{|\delta |}$,
corresponding to a zero {\em effective} phase mismatch term $(q^{2}+\delta )$, 
provided $\delta <0$. The total average number of atoms $\langle \hat{N}
(\tau )\rangle =\sum_{q}\left\langle \hat{a}_{q}^{\dagger }(\tau )\hat{a}
_{q}(\tau )\right\rangle $ is given by 
\begin{equation}
\langle \hat{N}(\tau )\rangle =\sum\nolimits_{q}(g/g_{q})^{2}\sinh
^{2}\left( g_{q}\tau \right) ,
\end{equation}
which grows exponentially with $\tau $.

To analyze correlations and relative number squeezing, we define particle
number operators $\hat{N}_{-}(\tau )$ and $\hat{N}_{+}(\tau )$ containing
only negative or positive momentum components, respectively: $\hat{N}
_{-(+)}(\tau )=\sum_{q<0(q>0)}\hat{a}_{q}^{\dagger }(\tau )\hat{a}_{q}(\tau
) $. We next consider the normalized variance $V(\tau )$ of the particle
number difference $[\hat{N}_{-}(\tau )-\hat{N}_{+}(\tau )]$, which -- in
normally ordered form -- is given by: 
\begin{equation}
V(\tau )=1+\langle :[\Delta (\hat{N}_{-}-\hat{N}_{+})]^{2}:\rangle /(\langle 
\hat{N}_{-}\rangle +\langle \hat{N}_{+}\rangle )  \label{V-define}
\end{equation}
where $\Delta \hat{X}\equiv \hat{X}-\langle \hat{X}\rangle $, and $V(\tau
)<1 $ implies squeezing of fluctuations below the coherent level -- which is
due to quantum correlations between $\hat{N}_{-}$ and $\hat{N}_{+}$.

Calculating $\langle :(\hat{N}_{\pm })^{2}:\rangle $ and $\langle \hat{N}_{-}
\hat{N}_{+}\rangle $, and using the fact that $B_{q}^{2}-|A_{q}|^{2}=-1$,
gives 
\begin{equation}
V(\tau )=1+\sum\nolimits_{q>0}B_{q}^{2}\left( B_{q}^{2}-|A_{q}|^{2}\right)
/\sum\nolimits_{q>0}B_{q}^{2}=0,
\end{equation}
implying perfect ($100\%$) squeezing in the particle number difference, at
least in this idealized calculation.

We now turn to the analysis of more realistic non-uniform case and including
the molecular field depletion, as described by the original coupled operator
equations for the fields, Eqs. (\ref{Heizneberg field-operator equations}).
These are solved numerically using equivalent stochastic ($c$-number)
differential equations in the positive-$P$ representation \cite{+P}, where
we additionally include a coupling to an atomic loss reservoir to describe
possible linear losses at a rate $\gamma $: 
\begin{eqnarray}
\frac{\partial \psi _{1}}{\partial \tau } &=&i\frac{\partial ^{2}\psi _{1}}{
\partial \xi ^{2}}-(\gamma +i\delta )\psi _{1}+\kappa \psi _{2}\psi _{1}^{+}+
\sqrt{\kappa \psi _{2}}\eta _{1\,\,},  \nonumber \\
\frac{\partial \psi _{1}^{+}}{\partial \tau } &=&-i\frac{\partial ^{2}\psi
_{1}^{+}}{\partial \xi ^{2}}-(\gamma -i\delta )\psi _{1}^{+}+\kappa \psi
_{2}^{+}\psi +\sqrt{\kappa \psi _{2}}\eta _{1\,\,}^{+},  \nonumber \\
\frac{\partial \psi _{2}}{\partial \tau } &=&\frac{i}{2}\frac{\partial
^{2}\psi _{2}}{\partial \xi ^{2}}-iv(\xi ,\tau )\psi _{2}-\frac{\kappa }{2}
\psi _{1}^{2}+\sqrt{-iu}\psi _{2}\eta _{2}\,\,,  \nonumber \\
\frac{\partial \psi _{2}^{+}}{\partial \tau } &=&-\frac{i}{2}\frac{\partial
^{2}\psi _{2}^{+}}{\partial \xi ^{2}}+iv(\xi ,\tau )\psi _{2}^{+}-\frac{
\kappa }{2}\psi _{1}^{+2}+\sqrt{iu}\psi _{2}^{+}\eta _{2}^{+}\,\,.
\label{Positive-P-eqs}
\end{eqnarray}
Here $\psi _{i}$ and $\psi _{i}^{+}$ are complex stochastic fields
corresponding respectively to the operators $\hat{\psi}_{i}$ and $\hat{\psi}
_{i}^{\dag }$, $v(\xi ,\tau )=[V_{2}(\xi d_{0})-V_{2}(\xi
_{0}d_{0})]t_{0}+u\psi _{2}\psi _{2}^{+}$ represents the effective
potential, and $\eta _{i}$ , $\eta _{i}^{+}$ are four real independent
delta-correlated Gaussian noise terms: $\left\langle \eta _{i}(\xi ,\tau
)\eta _{j}(\xi ^{\prime },\tau ^{\prime })\right\rangle =\langle \eta
_{i}^{+}(\xi ,\tau )\eta _{j}^{+}(\xi ^{\prime },\tau ^{\prime })\rangle
=\delta _{ij}\delta (\xi -\xi ^{\prime })\delta (\tau -\tau ^{\prime })$.

We consider molecules as an initial coherent field corresponding to
Thomas-Fermi inverted parabola for the molecular density. In this case,
assuming that $\psi _{2}(\xi ,0)$ is real, we have $\psi _{2}(\xi ,0)=\sqrt{
d_{0}n_{2}(\xi )}$, where $n_{2}(x)=n_{2}(0)\left[ 1-(\xi /\xi _{0})^{2}
\right] \theta (\xi _{0}-|\xi |)$. The molecular condensate axial
half-length is denoted via $\xi _{0}$ which is determined by the trap
geometry, and we assume repulsive molecule-molecule interactions. The time
duration for the molecule-atom conversion is controlled via $\kappa (\tau
)=\kappa _{0}\theta (\tau _{1}-\tau )$, so that $\kappa (\tau )=0$ for $\tau
>\tau _{1}$. Once the dissociation is stopped, we continue the evolution of
the resulting atomic field in free space to allow spatial separation of the
modes with positive and negative $q$ values.

For spatially separated components, we can introduce a pair of particle
number operators: 
\vspace{-0.25cm}
\begin{equation}
\hat{N}_{-(+)}(\tau )=\int_{-l/2(0)}^{0(l/2)}\hat{\psi}_{1}^{\dag }(\xi
,\tau )\hat{\psi}_{1}(\xi ,\tau )d\xi .  \label{N-plus-minus-space}
\vspace{-0.25cm}
\end{equation}
Next, we define the normalized variance $V(\tau )$ of the particle number
difference $[\hat{N}_{-}(\tau )-\hat{N}_{+}(\tau )]$ as before [see Eq. 
(\ref{V-define})], and evaluate the average values numerically using the
positive-$P$ Eqs. (\ref{Positive-P-eqs}) and the standard correspondence
between the normally ordered operator averages and the $c$-number stochastic
averages \cite{+P}.

\begin{figure}
\vspace{-0.7cm}
\centering \resizebox*{7.6cm}{!}{\includegraphics{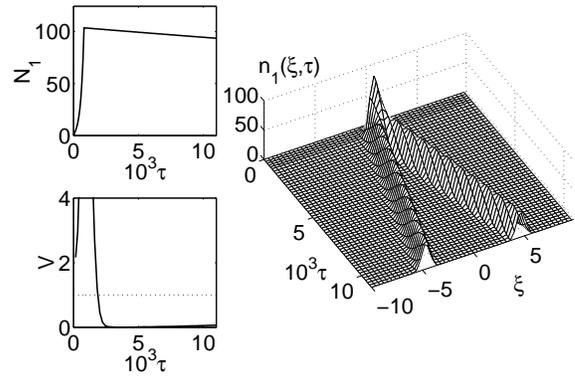} }
\vspace{-0.3cm}
\caption{
Atomic density $n_{1}(\protect\xi ,\protect\tau )$, total number of
atoms $N_{1}(\protect\tau )$, and the variance $V(\protect\tau )$
, for 40000 trajectory averages and for: $\protect\delta =-5.017\times 10^{4}$
, $u=1.8$, $\protect\kappa_{0}=84$, and $\protect\gamma =5$ . The
dimensionless dissociation time in this simulation is $\protect\tau 
_{1}=8\times 10^{-4}$. With the choice of the length scale $d_{0}=30$ $
\protect\mu $m and the mass of a $^{87}$Rb atom, we obtain the time scale $
t_{0}=2.446$ s, so that $\protect\tau =0.011$ corresponds to the total time
window $t\simeq 27$ ms, while the dissociation time $\protect\tau _{1}$
scales to $t_{1}\simeq 2$ ms. 
}
\label{Fig_1}
\vspace{-0.5cm}
\end{figure}

The results for $V(\tau )$, the total number of atoms in the two beams $
N_{1}(\tau )=\langle \hat{N}_{-}(\tau )\rangle +\langle \hat{N}_{+}(\tau
)\rangle $, and the density distribution $n_{1}(\xi ,\tau )=\langle 
\hat{\psi}_{1}^{\dag }(\xi ,\tau )\hat{\psi}_{1}(\xi ,\tau )\rangle $ are
represented in Fig. \ref{Fig_1}, for parameters that are reasonable in current
experiments and including a linear atomic loss term as discussed below.

The initial growth of the fluctuations in the atom number difference, during
the time interval when the atom-molecule coupling is switched on, is due to
the fact that the quantities $\langle N_{-}(\tau )\rangle $ and $\langle
N_{+}(\tau )\rangle $ each include atoms traveling in opposite directions
inside the molecular BEC region. The density distribution in coordinate
space at this stage is single peaked, due to the fact that it contains
amplified contributions from both momentum components. The fundamental
correlation of opposite momentum components is therefore not visible in the
atomic density initially. Once, however, the interaction is switched off and
the correlated atom pairs fly apart without further parametric
amplification, there is a double-peaked distribution, and we see a rapid
reduction of the variance below the coherent level, $V(\tau )<1$.

The physical reason for the correlation is momentum conservation, which
requires that each emitted atom with $q>0$ be accompanied by a partner atom
having $q<0$. In order to conserve energy, this pairing only occurs for $
\delta <0$, which allows potential energy in the molecule to be converted to
atomic kinetic energy for selected modes with $q$ values around $q_{0}=\pm 
\sqrt{|\delta |}$.

As the scheme relies on conservation laws for its operation, it should be
insensitive to the exact mode-structure. Clearly, quasi-one dimensional
traps are preferable for reasons of directionality, but we expect similar
results even if there is no transverse trap, provided that the molecules are
confined in a high aspect-ratio (cigar-shaped) trap to allow gain-guiding of
the atoms.

Investigations of losses show that these have a minimal effect on squeezing,
provided the atoms that are lost are only a fraction of the atoms produced.
For example, including an atomic loss term at a rate $\gamma /2=10$,
corresponding to the loss of $\sim 10\%$ of atoms during the free evolution
time interval, gives $V\simeq 0.07$ at $\tau =0.011$ as shown in Fig. 1.
This corresponds to a rather high ($\sim 93\%$) degree of squeezing below
the coherent level. Additional effects may occur when there is an atomic
potential, or when there is a strong atom-molecule scattering, but this has
very little effect on squeezing when $|\delta |\gg |u_{a}|$, if $u_{a}$ is
the scaled effective atomic potential.

The parameter values of Fig. 1 are derived using a length scale of $d_{0}=30$
$\mu $m taken to be equal to the molecular condensate axial half-length $
R_{x}$, so that $R_{x}\equiv x_{0}=d_{0}$ and $\xi _{0}=1$. We assume that
the molecular BEC is formed in a highly elongated trap and at densities that
satisfy the conditions for the crossover from $3D$ to $1D$ (see, e.g., \cite
{1D_BEC}). Assuming that the molecule-molecule scattering length $a_{2}$
is of the same order of magnitude as the scattering length of $^{87}$Rb
atoms, we take $a_{2}=a_{1}=5.4$ nm. We also take the molecular BEC linear
density at the trap center $n_{2}(0)=3.7\times 10^{7}$ m$^{-1}$ and an
aspect-ratio of $100$. This implies that the condensate transverse radius is 
$R_{\perp }=0.3$ $\mu $m, so that the $3D$ peak density is about $3.27\times
10^{19}$ m$^{-3}$. The initial total number of molecules is $
N_{2}=1.48\times 10^{3}$. Using the value of $R_{\perp }$ to scale out the
transverse confinement ($A\rightarrow \pi R_{\perp }^{2}$), we can next
estimate the one dimensional values of $U_{ij}$ and $\chi _{0}$. The $1D$
value of $\chi _{0}$ is obtained according to $\chi _{0}=\chi ^{(3D)}/\sqrt{
\pi R_{\perp }^{2}}$, where we take $\chi ^{(3D)}=2\times 10^{-7}$ m$^{1/2}/$
s \cite{Superchemistry}. The molecular condensate half-length of $30$ $\mu $
m corresponds to the trap axial oscillation frequency $\omega _{x}/2\pi =4.1$
Hz, where $\omega _{x}=\sqrt{2\hbar U_{22}n_{2}(0)/(m_{2}x_{0}^{2})}$. We
choose $\Delta =-2.051\times 10^{4}$ s$^{-1}$, thus assuring that the the
atom-molecule $S$-wave scattering is negligible in this calculation since $
|\Delta |\gg |U_{12}|\langle \hat{\Psi}_{2}^{\dagger }\hat{\Psi}
_{2}^{{}}\rangle $, where $U_{12}$ is estimated using $a_{12}\simeq -9.25$
nm \cite{WFHRH2000}. The final values of the parameters $\delta $, $u$, and $
\kappa _{0}$ that we arrive at are specified in the figure caption, together
with the relevant time scales.

In summary, we have shown that photo-dissociation of a molecular BEC can
provide a simple yet robust scheme for quantum squeezing of relative number
fluctuations in two counter-propagating atomic beams. The effects of
molecular condensate trapping and depletion, molecular self-phase
modulation, and atomic absorption have all been included in our
calculations. Our method does require high efficiency atom counting
techniques \cite{Aspect}, which are currently the subject of intensive
activity.

Applications may emerge from the use of these quantum-entangled twin beams
to produce a single beam with a well defined particle number, which can be
achieved by a destructive measurement on the partner beam. While this should
be readily observable, even more subtle experiments may be feasible in
future, including possible demonstrations of Einstein-Podolsky-Rosen
correlations or Bell inequalities \cite{Reid} in matter-wave quadratures.
Such experiments would open the way to novel tests of quantum mechanics for
macroscopic numbers of massive particles.

The authors gratefully acknowledge the ARC for the support of this work, and
Professor A. Ben-Reuven for useful discussions.

\end{document}